\documentclass[conf]{new-aiaa}
\usepackage[utf8]{inputenc}

\usepackage{graphicx}
\usepackage{amsmath}
\usepackage[version=4]{mhchem}
\usepackage{siunitx}
\usepackage{longtable,tabularx}
\usepackage{soul}
\title{Digital Twin of a DC Brushless Electric Motor-Propeller
\\ System with Application to Drone Dynamics}

\author{Daniel J. Gauthier\footnote{Developer and Co-Founder, ResCon Technologies, LLC, 1275 Kinnear Road Suite 239, Columbus, Ohio, 43212}} 

\author{Nicholas Biederman\footnote{Software Engineer, ResCon Technologies, LLC, 1275 Kinnear Road Suite 239, Columbus, Ohio, 43212}}
\author{Brian Gyovai\footnote{Founder/CEO, ResCon Technologies, LLC, 1275 Kinnear Road Suite 239, Columbus, Ohio, 43212}} 
\affil{ResCon Technologies, LLC, Columbus, OH 43212, USA}

\author{Jay P. Wilhelm\footnote{Associate Professor, Mechanical Engineering, Ohio University, Athens, Ohio, 45701, Senior Member}}
\affil{Ohio University, Athens, OH 45701, USA}

\begin{document}

\maketitle

\begin{abstract}
A digital twin of a direct current brushless (BLDC) electric motor and propeller is developed for predicting the generated thrust when there is no motion of the system (static conditions).  The model accounts for the back electromotive force, the propeller drag force, and the finite response time arising from the electromagnet winding inductance and DC resistance. The model is compared to a textbook model of BLCD dynamics and to experimental measurements on a KDE Direct KDE2315XF-885/885 Kv motor with a 945 propeller and a Holybro electronic speed controller (ESC) driving an AIR 2216/880 Kv motor with a 1045 propeller.  These systems are typically found on Group 1 uncrewed quadcopters (drones).  Both the steady-state and transient dynamics depart substantially from linearized models found in the literature.  This study is a starting point for disentangling the dynamics of the motor and the change in propeller dynamics due to complex airflow conditions.

\end{abstract}



\section{Introduction} \label{Sec:Intro}

A digital twin provides an accurate computer model of a system that can be used to predict its behavior over a range of conditions \cite{kahlen_digital_2017}.  It is used, for example, to shorten the design cycle, as a tool for predictive maintenance in a deployed system, or to integrate the system into a larger system. Here, we focus on developing a digital twin of a direct current (DC) brushless (BLDC) motor and propeller combination driven by a typical electronic speed controller (ESC) found in a small (Group 1 or 2) Uncrewed Aerial System (UAS).  We develop a dynamic model of the thrust produced by the system as a function of the pulse-width modulation (PWM) signal input to the ESC under static conditions when the motor/propeller system is held at a fixed position.  

The purpose of this study is to develop a digital twin of the motor/propeller system without the complication of the changing propeller efficiency with incident airflow.  Previous studies combine both effects (see, for example, Ref.~\cite{Khan2013}, or only focus on the propeller characteristics in different airflows (see, for example, Refs.~\cite{Brandt2011,McCrink2017,Dantsker2022}).  Our contribution is to clearly identify effects of the coupled electro-mechanical system.   Our future work will generalize this model to the motor-propeller system moving in air, which is important for multi-copters flying in high wind \cite{Masse2018,Dantsker2022}, in the presence of ground-effect \cite{Wilhelm2018}, or during aggressive maneuvers \cite{Kaufmann2020}

For the class of sensorless ESCs found on typical multi-copters, we find that the thrust is a nonlinear function of the PWM signal, which is especially pronounced at lower PWM signals (slower motor rotation rate) and approximately linear for higher PWM signal.  We also find that the response to a step-change in the PWM signal is non-exponential and that is is important to include the electrical properties of the BLDC.  These characteristics are largely ignored in the drone control literature, but they are important for understanding UAS flight dynamics. 

In the next section, we discuss the basic operation of a DC brushless electric motor (BLDC) and a motor-propeller model described in a popular aircraft control and simulation textbook \cite{StevensBook} and one used in the PX4 open-source multi-copter flight control system \cite{PX4}.  We then introduce a electro-mechanical digital twin of the static motor-propeller model in Sec.~\ref{Sec:PhysicsModel},  compare the model predictions to experimental observations in Sec.~\ref{Sec:Experiment}, and conclude with implications of our results.


\section{Background} \label{Sec:Background}

A BLDC-propeller combination is found in a wide variety of electrified aircraft technologies because of their high thrust-to-weight ratio, the recent availability of high energy density batteries, and the development of solid-state electronics used for fast switching of the voltage applied to the motor electromagnet windings \cite{Microchip}.

Torque is generated from BLDC's by keeping the magnetic fields of the rotating part of the motor (rotor) and the stationary part (stator) misaligned.  Rotors on BLDC's typically include 2$N$ high-energy-density permanent magnets around an outer ring, which is connected to the shaft in the center of the motor.  The stator is in between the outer ring of magnets and the shaft and contains 2$M$ electromagnets.  The misalignment between the electromagnet and permanent magnet fields generates a torque that attempts to align the fields; as they approach alignment, the neighboring electromagnet is energized and the near-aligned one is de-energized, thus maintaining misalignment of the fields and the magnets \cite{Microchip} and producing a continuous torque.  A typical BLDC has $N,~M \ge$ 3.

The ESC uses solid-state switches to energize/de-energize the electromagnet windings based on two approaches:  sensored ESCs use real-time direct measurement of the shaft angle, typically Hall probes; or sensorless ESCs that estimate the angle by measuring the so-called back electromotive force (EMF) \cite{Microchip}.  Sensored ESCs are available using modern DShot-protocol ESCs operating in angular rate control mode, for example, but they are rarely used because of the increased cost of the motor sensor and the ESC.  Inexpensive ESCs used in small UASs tend to measure the back-EMF. The vast majority of small UAS motor controllers use sensorless ESC's because of cost, availability, and functionality.

The ESC is often controlled using a PWM signal: a square-wave voltage signal with a frequency in the 10's of kHz range with the duty cycle of the waveform controlled by the user, typically in the range between 50\% (0\% throttle) and 100\% (100\% throttle). This signal should not be confused with the modulated signal applied to electromagnet windings whose frequency is locked to the propeller rotation rate but whose duty cycle is related to the duty cycle of the user-input signal to the ESC. 

\subsection{Thrust Characteristics of BLDC-propeller Systems} \label{Sec:ForceIntro}

The force (thrust) generated by a BLDC-propeller system in the direction along the motor shaft is typically modeled as \cite{Masse2018}
\begin{equation}
F = k_t \omega^2, \label{Eq:thrust}
\end{equation}
where $k_t$ is the thrust coefficient in N s$^2$/rad$^2$ and $\omega$ is the angular velocity of the motor-propeller in rad/s. The thrust coefficient is a nonlinear function of $\omega$ and the linear velocity of the center of the propeller in the air, which we take as having a component parallel $\upsilon_{\parallel}$ and perpendicular $\upsilon_{\perp}$ to the motor shaft.  The velocity includes the free-stream and wind velocities \cite{Masse2018}.  

Most studies of the variation of $k_t$ with velocity are for fixed-wing aircraft where $\upsilon_{\perp}\sim 0$; the effects of the perpendicular velocity are usually treated as a separate force.  Following this approach, $k_t$ is a function of the advance ratio 
$J=2 \pi \upsilon_{\parallel}/\omega D$, where $D$ is the propeller diameter.  For multi-copter hover conditions with no wind or ground effect, $J\sim 0$ and $k_t$ is equal to its static value $k_{t0}$.

For inexpensive plastic propellers with diameters <12", $k_t$ decreases approximately linearly with $J$ up to values of $J \sim 0.4$, beyond which the drop off is more rapid until attaining the zero-thrust condition \cite{Brandt2011,McCrink2017}.  For larger propellers, $k_t$ is relatively constant for $J< \sim 0.1$, then drops off as for the smaller propellers \cite{Dantsker2022}.  As discussed above, we only consider the case when $J=0$ for which $k_t=k_{t0}$ is a constant to isolate the dynamics of the electro-mechanical system in static flight conditions.

\subsection{PX4 Motor Model} \label{Sec:PX4}

The motor model used in the open-source PX4 flight stack for a multi-copter has no time dependence and is described in the \textbf{Thrust Curve} section of the controller guide \cite{PX4Thrust}.  Here, the dimensionless throttle command $T = 2 (\mathrm{PWM}-0.5)$ (0-1) is related to the relative thrust through the relation
\begin{equation}
    \frac{F}{F_{max}} = \mathrm{THR\_MDL\_FAC} ~ T^2 + (1-\mathrm{THR\_MDL\_FAC})~ T, \label{Eq:PX4}
\end{equation}
where $F_{max}$ is the maximum force produced by the system at the maximum rotation rate, and $\mathrm{THR\_MDL\_FAC}$ is a dimensionless parameter in the range 0-1.  When $\mathrm{THR\_MDL\_FAC}$ = 0 (the default value), the thrust is linearly proportional to the throttle command.  When $\mathrm{THR\_MDL\_FAC}$ = 1, the thrust depends on the square of the throttle command, which is appropriate for an ESC that controls the angular velocity of the motor with the speed proportional to the throttle command.  For ESCs that do not control the angular velocity, the guide suggests taking $\mathrm{THR\_MDL\_FAC}$ in the range of 0.3 to 1.  

As we show in Secs.~\ref{Sec:SSTheory} and \ref{Sec:SSResults} below, this model does not agree with the steady-state predictions of the physics-based model of the motor-propeller system and our experimental observations using a force stand.  It also ignores the finite response time of the motor, although an effective exponential time constant can be included using a low-pass filter in the flight stack control loop.





\subsection{Simple BLCD motor model} \label{Sec:StevensModel}

Stevens \textit{et al.} \cite{PX4Thrust} present a medium-fidelity model of a quadcopter that proposes a model of the ESC that controls $\omega$ to achieve a desired angular velocity $\omega_d$ and a first-order lag for the tracking of the ESC control loop. In this case, the dynamics of the ESC-BLDC-propeller system are described by
\begin{equation}
\frac{d\omega}{dt} = \frac{1}{\tau_{ESC}}(\omega_d - \omega), \label{Eq:Stevens}
\end{equation}
where $\tau_{ESC}$ is the response time of the ESC control loop.

\section{Physics-Based Static BLDC-Propeller Model} \label{Sec:PhysicsModel}

Our model follows that presented in Ref.~\cite{Masse2018} generalized to account for the operation of the ESC.  Here, we focus only on sensorless ESCs, where the duty cycle of the user-input signal is related to the duty cycle of the signals driving the electromagnet coils.  This results in an effective voltage of \cite{Microchip}
\begin{equation}
V = V_{batt} T = 2 V_{batt} (\mathrm{PWM}-0.5) \label{Eq:ESC}
\end{equation}
applied to the electromagnet windings, where $V_{batt}$ is the battery voltage.  This relation breaks down at low voltages (small $T$) because the back EMF is too small to measure at  low $\omega$.

\begin{figure}[hb]
	\centering
	\includegraphics[width=0.5\linewidth]{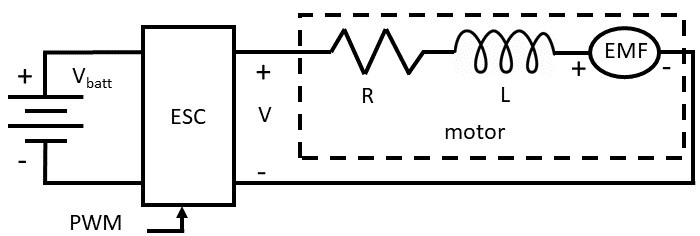}
	\caption{Equivalent model of the motor system.}
	\label{fig:circuit}
\end{figure}

The presence of multiple electromagnet windings makes the BLCD electrical circuit somewhat complicated, but it is known \cite{Microchip} that the simpler equivalent circuit shown in Fig.~\ref{fig:circuit} gives accurate results. It consists of the ESC, which reduces the battery voltage $V_{batt}$ according to Eq.~\ref{Eq:ESC}, a series inductor of inductance $L$, a resistor of resistance $R$, and a voltage source representing the back EMF, which reduces the voltage applied to the BLCD terminals.  Here, the back EMF has magnitude $k_e \omega$, where $k_e$ is the counter EMF constant in units of V s/rad.  It can be estimated, through appropriate unit conversion, to the inverse of motor "Kv" constant typically specified by the vendor; that is, $k_e \sim 1/\mathrm{Kv}$.  In the next section, we give an expression for $k_e$ that can be obtained from the motor/propeller steady-state thrust data.

The coupled electro-mechanical model is found by applying Kirchhoff's Laws to the circuit and Newton's Laws of rotational motion to the propeller-motor combination \cite{Masse2018}. They are coupled because the back EMF depends on $\omega$ and the torque applied to the motor shaft is proportional to the current $i$ flowing through the circuit. This analysis results in a coupled electro-mechancial model given by
\begin{eqnarray}
    L \frac{di}{dt} = V - k_e \omega - Ri, \label{Eq:i} \\
    J_m \frac{d\omega}{dt} = k_m i - b_m \omega - k_q \omega^2, \label{Eq:omega}
\end{eqnarray}
where $J_m$ is the propeller-motor moment of inertia, $k_m = k_e$ is the motor torque constant, $b_m$ is the viscous friction constant, and $k_q$ is the propeller drag constant.  In most studies, the propeller drag moment exceeds the viscous drag moment and hence it is typical to take $b_m$=0, which we do so here.  However, it is straightforward to generalize our results to include the viscous drag moment. Table \ref{Tab:param} gathers these constants for easy reference. 

\begin{table}[ht]
	\caption{Model Parameters}
	\begin{center}
		\begin{tabular}{|c|c|c|c|c|c|}
			\hline
			\textbf{Symbol}&\textbf{Parameter}&\textbf{Units} & \textbf{Symbol}&\textbf{Parameter}&\textbf{Units}\\
			\hline
                $k_t$ & thrust coefficient & N s$^2$/rad$^2$ & $k_{t0}$ & static thrust coefficient & N s$^2$/rad$^2$ \\
                
                $k_q$ & propeller drag coefficient & N m s$^2$/rad$^2$ & $J_m$ & moment of inertia & kg m$^2$ \\
                
                $J$ & propeller advance ratio & dimensionless & $k_m$ & torque constant & N m/A \\
                
                $b_m$ & viscous friction constant & N m s/rad & $k_e=k_m$ & counter EMF constant & V s/rad \\
                
                $L$ & inductance & H & $R$ & resistance & $\Omega$ \\
                
                $\tau=L/R$ & motor time constant & s & $V_{batt}$ & battery voltage & V \\
                
                $T$ & throttle command & dimensionless & PWM & pulse width mod. signal & dimensionless \\
                
                $\omega_d$ & desired angular velocity & rad/s & $\tau_{ESC}$ & ESC response time & s \\
                
                $\alpha$ & physics motor parameter & rad/s & $\beta$ & physics motor parameter & rad$^2$/s$^s$ \\
                
                $\omega_{max}$ & max. angular velocity & rad/s & & & \\
			\hline
		\end{tabular}
		\label{Tab:param}
	\end{center}
\end{table}

The model is noteworthy in that the dynamics of the angular velocity (Eq.~\ref{Eq:i}) has a quadratic dependence on $\omega$ due to the propeller drag and it is not described as a first-order equation given by Eq.~\ref{Eq:Stevens} due to its coupling to the motor current.  These characteristics have an impact on both the steady-state and time response of the motor to changes in the input PWM signal.

\subsection{Steady-state thrust} \label{Sec:SSTheory}

The steady-state behavior of the BLDC is found by setting the derivatives in Eqs.~\ref{Eq:i} and \ref{Eq:omega} to zero and solving for the steady-state (mnemonic $ss$) current and angular velocity using Eq.~\ref{Eq:ESC} for the thrust.  We find that 
\begin{eqnarray}
 \omega_{ss} &=& - \alpha + \sqrt{\alpha^2+\beta T}, \label{Eq:omegaSS} \\
 i_{ss} &=& \frac{V_{batt} T - k_e \omega_{ss}}{R}, \label{Eq:iss} \\
 \alpha &=& \frac{k_m k_e}{2 k_q R}, \label{Eq:alpha} \\
 \beta &=& \frac{k_m V_{batt}}{k_q R}. \label{Eq:beta} 
\end{eqnarray}
The parameters $\alpha$ and $\beta$ are not independent because $\omega_{ss} = \omega_{max}$ for $T=1$, where $\omega_{max}$ is the maximum BLDC angular velocity.  Using this fact, we find that 
\begin{equation}
\beta = (\omega_{max}+\alpha)^2 - \alpha^2 = \omega_{max}^2+2\alpha\omega_{max}. \label{Eq:beta-alpha}
\end{equation}
Thus,
\begin{equation}
\omega_{ss} = -\alpha+\sqrt{\alpha^2+(\omega_{max}^2+2\alpha\omega_{max})T}. \label{Eq:omega_only_alpha}
\end{equation}
This observation is important because it implies that there is only a single parameter ($\alpha$) that specifies the steady-state motor dynamics once $\omega_{max}$ is known.

Once $\alpha$ is obtained from a fit to Eq.~\ref{Eq:omega_only_alpha} combined with Eq.~\ref{Eq:thrust} to predict thrust as a function of $T$, we can obtain $\beta$ using Eq.~\ref{Eq:beta-alpha}.  Then, other model parameters can be obtained by considering the steady-state solutions with $T=1$. We find the counter EMF constant and torque constant through the relation
\begin{equation}
k_e = k_m = 2V_{batt}\frac{\alpha}{\beta}. \label{Eq:ke}
\end{equation}
From the steady-state solution to the current equation, we find the motor resistance
\begin{equation}
R = \frac{V_{batt}-k_e \omega_{max}}{i_{max}},
\end{equation}
and from the angular velocity equation we obtain the propeller drag coefficient
\begin{equation}
k_q = \frac{k_m i_{max}}{\omega_{max}^2}. \label{Eq:kq}
\end{equation}

\subsubsection{Limiting behaviors} \label{Sec:SSTheoryLimits}

For low enough angular velocities, we have that $\beta T/\alpha^2 \ll 1$ so that a Taylor series expansion of the square root in Eq.~\ref{Eq:omegaSS} is possible.  In this limit, we find that 
\begin{eqnarray}
\omega_{ss} &\approx& \frac{\beta}{2 \alpha} T, \label{Eq:omegaSSapprox} \\
F_{ss} &\approx& k_t \frac{\beta^2}{4 \alpha^2} T^2~~ (T \ll 1), \label{Eq:F_ss_approx}
\end{eqnarray}
and hence the thrust scales quadratically with the throttle for low throttle.  For larger values of $T$, $\omega_{ss}$ depends on the the size of $\alpha$ relative to $\omega_{max}$.  For $\alpha/\omega_{max} \ll 1$,
\begin{eqnarray}
    \omega_{ss} &\approx& \omega_{max} \sqrt{T}, \\
    F_{ss} &\approx& k_t \omega_{max}^2 T,  \label{Eq:F_ss_approx2}    
\end{eqnarray}
demonstrating that the BLDC thrust is linear with the throttle.

\begin{figure} [h]
	\centering
	\includegraphics[width=0.5\linewidth]{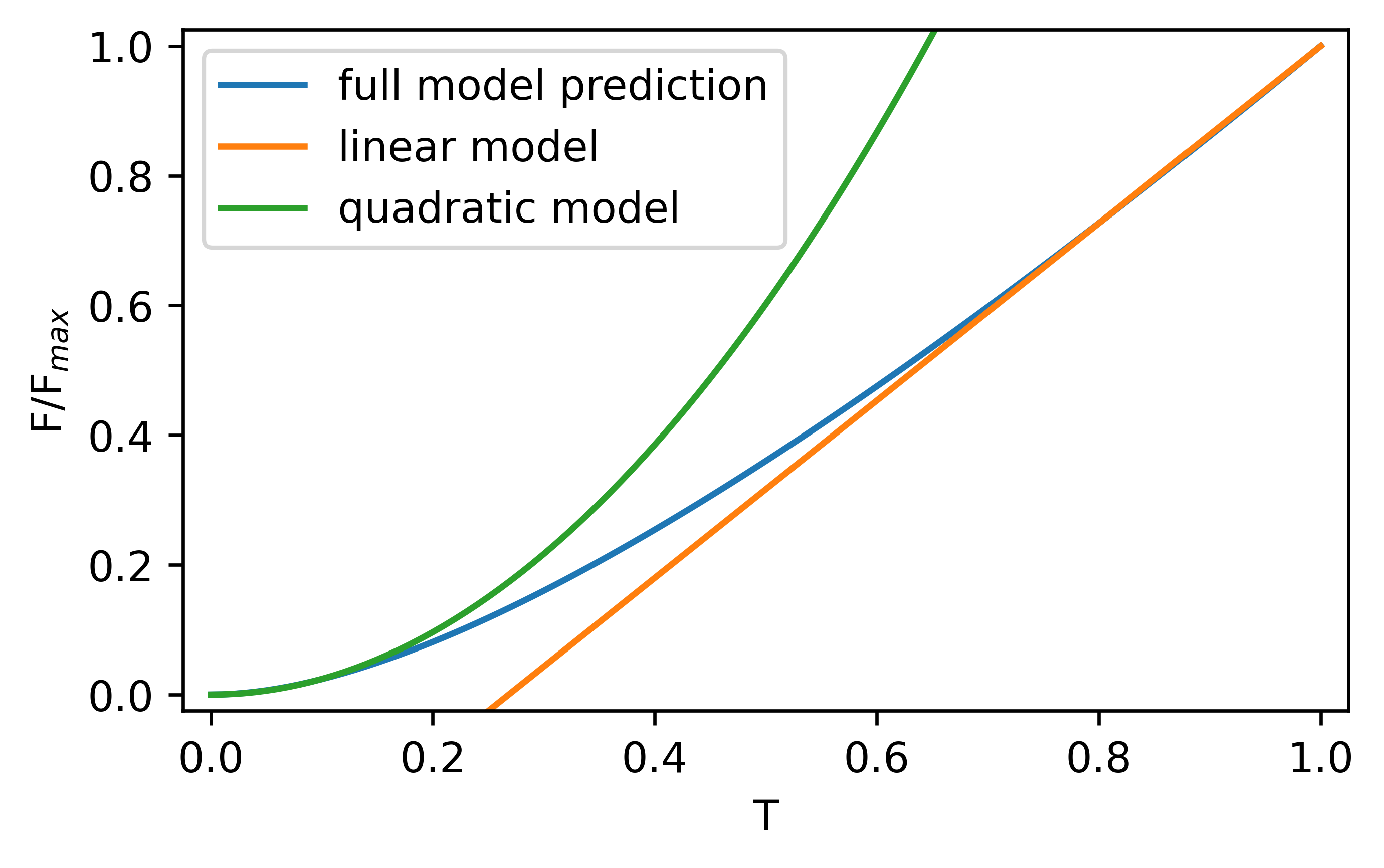}
	\caption{Scaling of the force. Physics-based ESC-BLDC-propeller model (blue line) compared to a quadratic scaling (green line) and linear scaling (orange line) with $\alpha=800$ and $\omega_{max}=1144$ rad/s.}
	\label{fig:dimensionlesstheory}
\end{figure}

For $\alpha \sim \omega_{max}$, which we hypothesize is the case for a typical BLDC-propeller system, the dependence of $F$ on $T$ is more complex and smoothly connects these two limiting behaviors.  As seen in Fig.~\ref{fig:dimensionlesstheory}, the dimensionless force scales quadratically for small $T$ as expected based on Eq.~\ref{Eq:F_ss_approx} and going over to linear scaling for $T\rightarrow 1$ as expected based on Eq.~\ref{Eq:F_ss_approx2}.

\subsection{Dimensionless Dynamic Model of the BLDC-Propeller System} \label{Sec:Dimensionless-model}

The steady-state quantities identified in the previous section can be used to put the differential equations of the model (Eqs.~\ref{Eq:i} and \ref{Eq:omega}) in dimensionless form, which will allow us to identify the remaining unknown parameters of the model.  To this end, we define dimensionless quantities $\tilde{i} = i/i_{max}$, $\tilde{\omega} = \omega/\omega_{max}$, $\tilde{t} = t/\tau$, $\tilde{\alpha} = \alpha/\omega_{max}$, and $\tilde{\beta} = \beta/\omega_{max}^2$, substitute these definitions into the differential equations, and use the parameters defined in the previous section to find
\begin{eqnarray}
\frac{d\tilde{i}}{d\tilde{t}} &=& \tilde{\beta}~T - \tilde{\alpha} \tilde{\omega} - \tilde{i} , \label{Eq:dim-i} \\
\frac{d\tilde{\omega}}{d\tilde{t}} &=& \left( \frac{L k_m i_{max}}{J_m \omega_{max}} \right)(\tilde{i} - \tilde{\omega}^2), \label{Eq:dim-omega}
\end{eqnarray}
with $b_m = 0$.  We see that the two remaining unknown model parameters, $L$ and $J_m$, do not appear independently but as the ratio $L/J_m$.  Thus, to fit dynamic data as we do in Sec.~\ref{Sec:DynamicResults} below, only a single ratio needs to be adjusted.

\subsection{Dynamic thrust} \label{Sec:DynamicThrustTheory}

The coupled electro-mechanics dynamical model given by Eqs.~\ref{Eq:i} and \ref{Eq:omega} can be cast as a Riccati equation and then transformed to a linear second-order differential equation.  However, the form of the solution is algebraically complicated and we do not obtain substantial insight from this approach.  On the other hand, we observe that the BLDC current has a first-order decay with time constant $\tau = L/R$.  But, due to coupling to $\omega$, the time course of the coupled system is non-exponential due to the propeller drag moment, which depends quadratically on $\omega$.

\begin{figure} [h]
	\centering
	\includegraphics[width=0.5\linewidth]{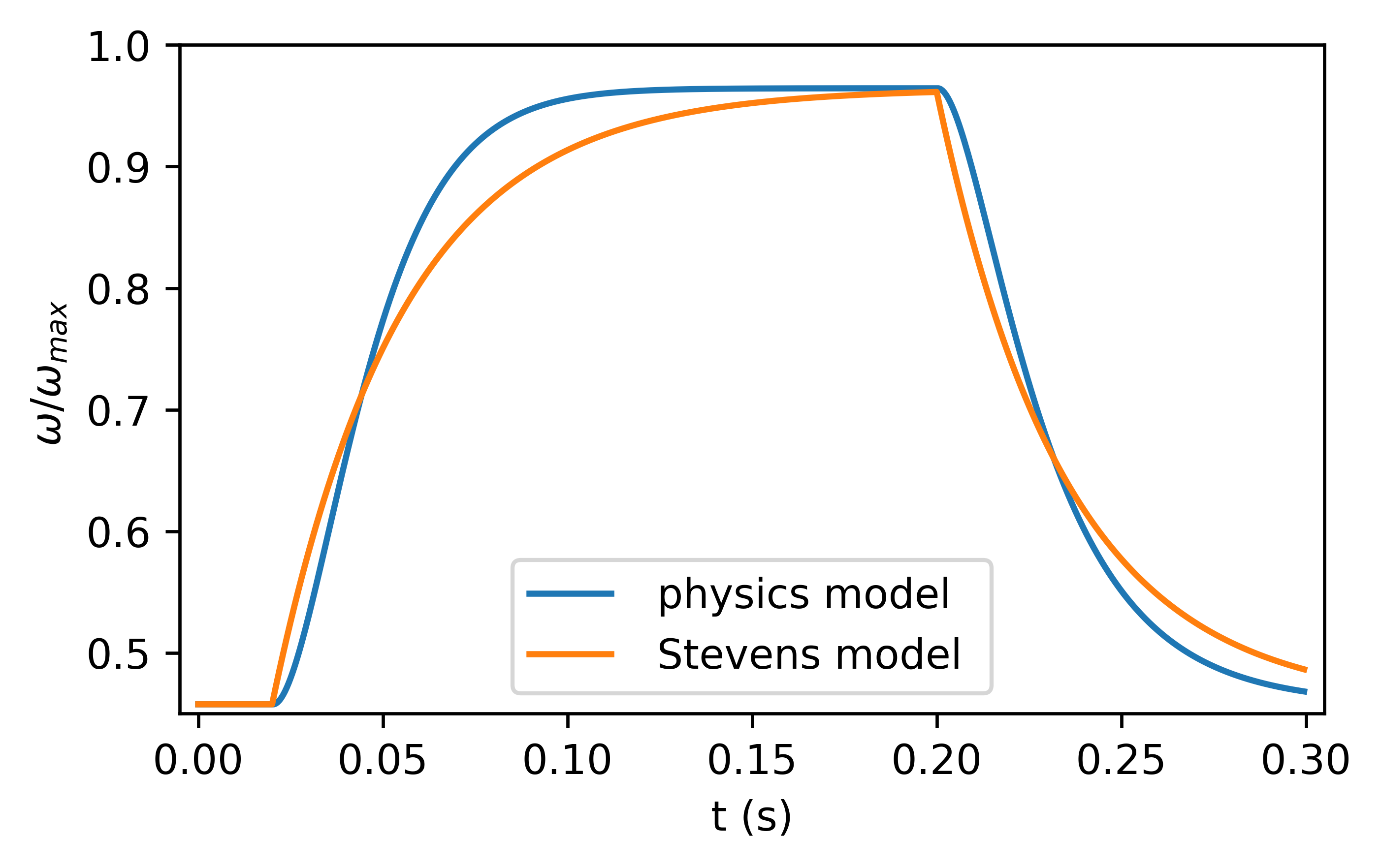}
	\caption{Motor dynamics. Temporal evolution of the ESC-BLDC-propeller dimensionless angular rate in response to a step change of $T$ for the physics based model (blue line) compared to the Eq.~\ref{Eq:Stevens} (orange line).  Here, $\alpha$=800 rad/s, $\omega_{max}$=1144 rad/s, $V_{batt} = 16$ V, $R$ = 0.33 $\Omega$, $L/J_m$ = 300 H/kg m$^2$,  $k_q = 1.94 \times 10^{-7}$ N m s$^2$/rad$^2$, and $k_e = 1.08 \times 10^{-2}$ V s/rad.}
	\label{fig:dynamicstheory}
\end{figure}

Figure \ref{fig:dynamicstheory} compares the physics-based model and the model proposed by Stevens \textit{et al.} (Eq.~\ref{Eq:Stevens}, where we have adjusted $\tau_{ESC} = 35$ ms so that the predictions agree at the half-height location. The parameters used to generate Fig.~\ref{fig:DynamicResults} are typical for a small UAS motor system discussed in Sec.~\ref{Sec:Experiment} below.  We see that the physics model reaches the equilibrium value more quickly than expected for a first-order lag system.  The disagreement between the models is substantial on the approach to equilibrium. 

Another observation is that $\tau_{ESC}$ is substantially longer than the time constant of the electrical circuit alone, which is given by $\tau = L/R$ = 9.0 ms for this example.  The motor angular velocity has no linear damping (we assume the viscous drag is zero in this simulation), so one might have guessed that $\omega$ just follows the change in $i$ and hence the time scale should be much faster.  This is not the case and points out the importance of analyzing the coupled dynamics of the electro-mechanical system.

\section{Experimental measurements and model comparison} \label{Sec:Experiment}

Many commercial BLDC vendors provide datasheets of the steady-state ESC-BLDC-propeller characteristics for different voltages and static flight conditions.  We use data provided by KDE Direct for the KDE2315XF-885 (885 Kv) motor with a 9' $\times$ 4.5" pitch propeller at 16 V (slightly higher than a 4S).  They do not specify the protocol used by the ESC, but the data shown below clearly demonstrates that the angular velocity is not controlled and hence we assume that a sensorless ESC of the type considered here was used in the measurements.

Less data is available on the dynamical response of ESC-BLDC-propeller systems from the manufacturer.  We used a force cell to measure the thrust of a Holybro ESC driving a AIR 2216/880 KV motor with a 1045 (10" diameter, 4.5" pitch) propeller. The parameters of this system are  similar to the KDE Direct system considered for the static measurement. 

The thrust measurements allow us to quantify the non-linear thrust and time response to commanded inputs. The test stand used, a Tyto series 1585, contains a load cell, measurement electronics, and sensorless ESC outputs. Their software allows for automatic generation of the thrust as a function of the angular velocity and precise control over commands for the ESC to collect time response curves. We collect several thrust curves and commanded angular velocity changes from stopped to full speed with the motor and propeller mounted horizontally.

\subsection{Steady-state results} \label{Sec:SSResults}

\begin{figure} [h]
	\centering
	\includegraphics[width=0.5\linewidth]{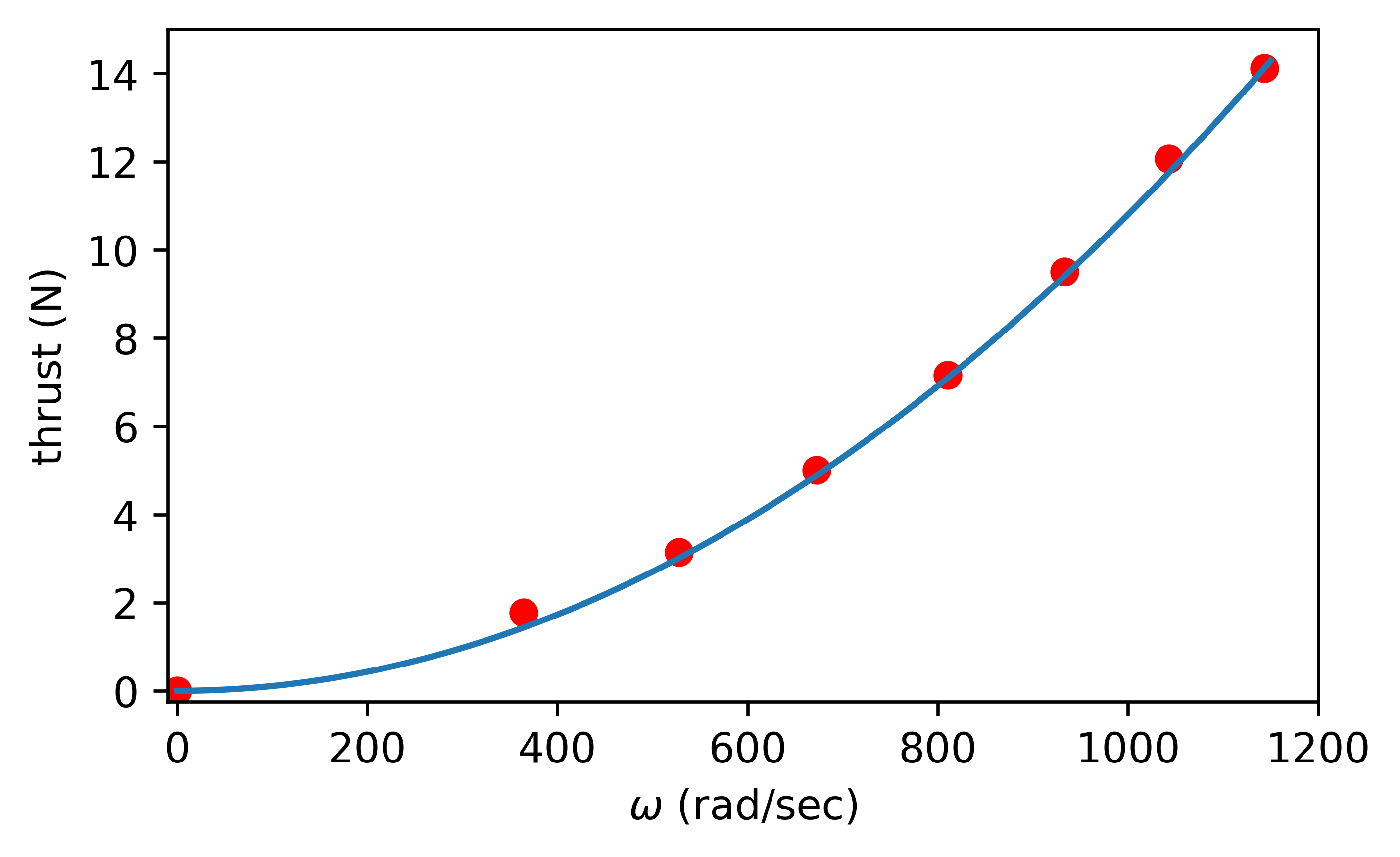}
	\caption{Steady-state thrust characteristics. Dependence of the thrust on angular velocity for the KDE Direct system from which we determine $k_{t0} = 1.08\times10^{-5}$. }
	\label{fig:KDEDirectThrustVSOmega}
\end{figure}

The first step in fitting the physics model parameters to experimental observations is to study the thrust as a function of angular velocity, shown in Fig.~\ref{fig:KDEDirectThrustVSOmega} for the KDE Directo system.  The solid line is a fit Eq.~\ref{Eq:thrust}, which identifies $k_{t0}=1.08 \times 10^{-5}$.  The fit is good over the entire range, indicating that the propeller is well matched to the motor.

\begin{figure} [bh]
	\centering
	\includegraphics[width=0.5\linewidth]{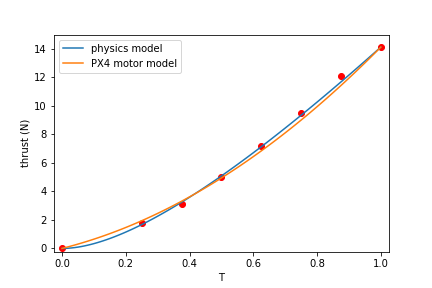}
	\caption{Transformed steady-state thrust characteristics. Dependence of the thrust on dimensionless throttle for the KDE Direct system from which we determine $\alpha = 800$. }
	\label{fig:KDEDirectThrustVSThrottle}
\end{figure}

The next step is to fit the physics model parameter for the thrust as a function of throttle, shown in Fig.~\ref{fig:KDEDirectThrustVSThrottle}.  We find the best fit is for $\alpha = 800$ rad/s, where it is seen that the fit is good over the entire throttle range.  We then use the fit $\alpha$ to obtain several other model parameters as mentioned above.

We also fit the data to the PX4 motor model (Eq.~\ref{Eq:PX4}, with the best fit obtained for THR\_MDL\_FAC = 0.6.  While the error between the fit of the PX4 model and the data is not large, the incorrect scaling with $T$ at small values and values near 1, discussed in Sec.~\ref{Sec:SSTheory} above, is evident.

\subsection{Dynamic results} \label{Sec:DynamicResults}

\begin{figure}
	\centering
	\includegraphics[width=0.5\linewidth]{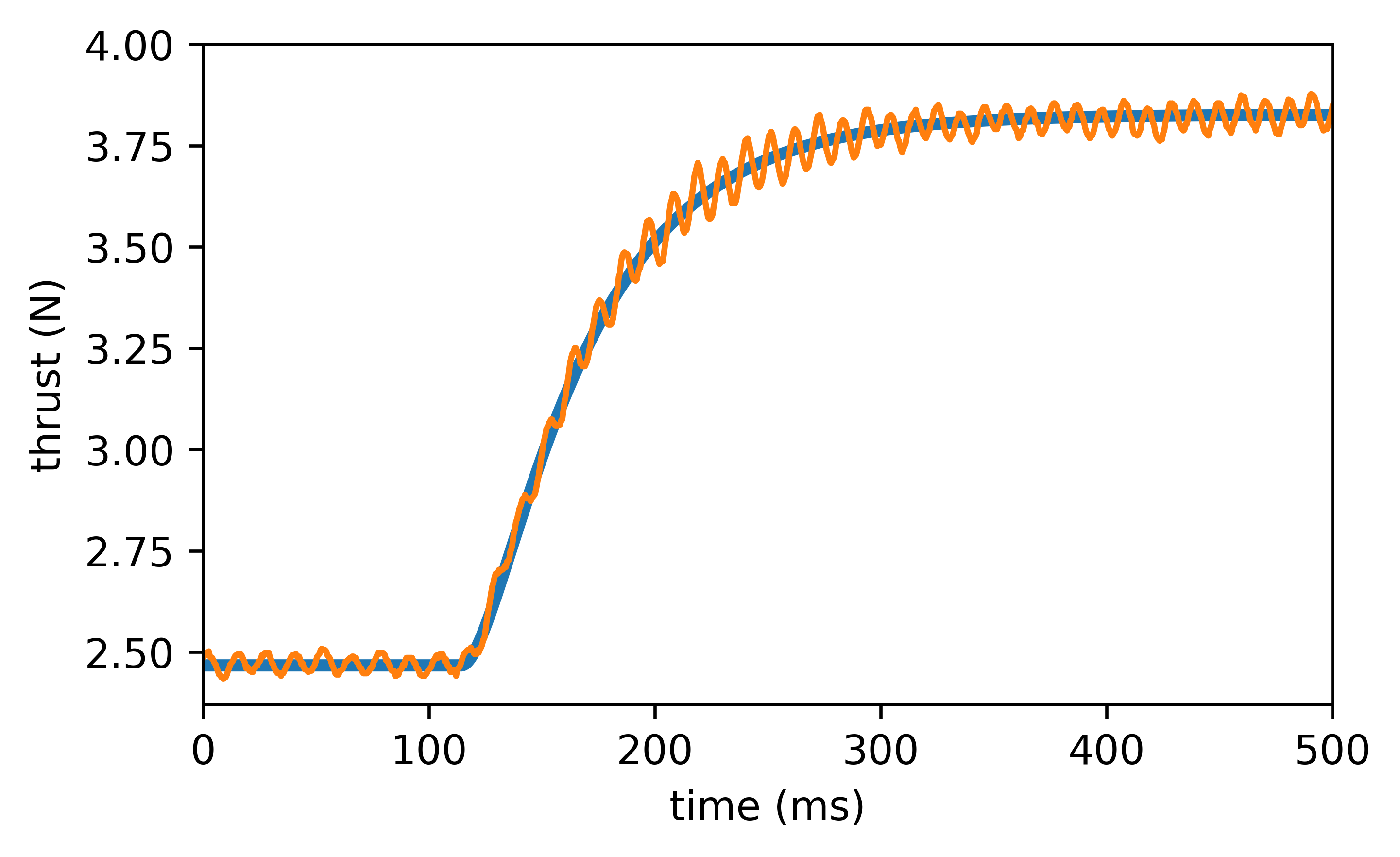}
	\caption{Observed transient motor dynamics.  The Holybro model parameters are: $\alpha$=800 rad/s, $\omega_{max}$=1144 rad/s at $V_batt = 16$ V used in the Holybro steady-state thrust measurements, $V_{batt} = 14.8$ V in the experiment recording the transient behavior, $R$ = 0.35 $\Omega$, $L/J_m$ = 172 H/kg m$^2$,  $k_q = 1.2 \times 10^{-7}$ N m s$^2$/rad$^2$, and $k_e = 8.16 \times 10^{-3}$ V s/rad.}
	\label{fig:DynamicResults}
\end{figure}

For the Holybro system, we observe the dynamics shown in Fig.~\ref{fig:DynamicResults} for a step change in the throttle from $T=0.34$ to $T=0.45$.  We see that the time to attain steady state is many 10's of milliseconds. 

We follow a similar procedure described above to find the model parameters for the Holybro system and adjust for the different battery voltage (the Holybro-provided data used $V_{batt}=16$ V, whereas we use $V_{batt}=14.8$ V in the experiment).  We then adjust $L/J_m$ to fit the temporal evolution of the observed and predicted behaviors, where we see the agreement is good.   

\section{Determining Model Parameters for a New BLDC-Propesser System}

The following procedure describes how to find the parameters for our model for a new BLDC-propeller system:
\begin{enumerate}
  \item From the manufacturer datasheet of the motor/propeller system, which usually gives steady-state behaviors, identify $V_{batt}$, $\omega_{max}$, $F_{max}$, and $i_{max}$.
  \item Plot thrust as a function of angular velocity as done in Fig.~\ref{fig:KDEDirectThrustVSOmega} and fit to Eq.~\ref{Eq:thrust} to find $k_t$.
  \item Plot the steady-state thrust as a function of the throttle command (Fig.~\ref{fig:KDEDirectThrustVSThrottle}) and fit to Eq.~\ref{Eq:omega_only_alpha} to find $\alpha$.
  \item Use Eq.~\ref{Eq:beta-alpha} to find $\beta$.
  \item Use Eqs.~\ref{Eq:ke}-\ref{Eq:kq} to find $k_e$, $k_m$, $R$, and $k_q$.
  \item To use this model to predict the steady-state behavior for a different battery voltage, use Eq.~\ref{Eq:beta} to find the new value of $\beta$.
  \item Collect dynamic data of the thrust for a step change in the throttle, preferably for a step change in the region where the thrust is a nonlinear function of the throttle (\textit{e.g.}, $T \sim$ 0.5).  Fit the data to model prediction by adjusting the ratio $L/J_m$. 
\end{enumerate}

\section{Discussion} \label{Sec:Discussion}

We present a physics-based model of an ESC-BLDC-propeller model consisting of coupled differential equations for the motor current and propeller angular velocity.  The coupling gives rise to a nonlinear thrust as a function of the throttle, which has the opposite scaling of the motor model currently used in the PX4 flight-control system.  Also, the dynamics are described by a double-lag solution as opposed to a single-lag solution often found in the literature.  The controller used in the flight stack can likely be improved using our model for systems that use an ESC that senses the back-EMF.

These results will give new insights for tuning flight control systems and will inform the design of artificial neural network - based controllers.  In the future, we will generalize the model to account for the system moving in air, which will be important for understanding how to design flight controllers for windy situations or aggressive flight maneuvers.

\section*{Acknowledgements}
Experimental results for this project were funded in-part by the United States Air Force AFRL/SBRK under Contract No. FA864921P0087 and the State of Ohio through the Ohio Third Frontier Technology Validation and Startup Fund (TVSF) grant program.

\bibliography{refs}

\end{document}